\documentclass[10pt,reprint,amsmath,amssymb,aps,pra,showpacs,longbibliography,superscriptaddress]{revtex4-1}
\usepackage[latin9]{inputenc}
\usepackage{amsmath}
\usepackage{amssymb}
\usepackage{graphicx}
\usepackage{esint}

\makeatletter

\usepackage{amsfonts}
\usepackage{graphics}

\makeatother

\begin{document}
\title{Spectral function of Fermi polarons at finite temperature from a self-consistent
many-body $T$-matrix approach in real frequency}
\author{Hui Hu}
\email{hhu@swin.edu.au}

\affiliation{Centre for Quantum Technology Theory, Swinburne University of Technology,
Melbourne 3122, Australia}
\author{Xia-Ji Liu}
\email{xiajiliu@swin.edu.au}

\affiliation{Centre for Quantum Technology Theory, Swinburne University of Technology,
Melbourne 3122, Australia}
\date{\today}
\begin{abstract}
We theoretically examine the finite-temperature spectral function
of Fermi polarons in three dimensions, by using a self-consistent
many-body $T$-matrix theory in real frequency. In comparison with
the previous results from a non-self-consistent many-body $T$-matrix
approach, we show that the treatment of self-consistency in the impurity
Green function leads to notable changes in almost all the dynamical
quantities, including the vertex function, impurity self-energy and
spectral function. Eventually, it gives rise to quantitatively different
predictions for the measurable radio-frequency spectrum and Raman
spectrum at finite temperature. Using the recent spectroscopic measurements
as a benchmark, we find that the self-consistent many-body $T$-matrix
theory somehow provides a better explanation for the experimental
data. The notable difference in the predictions from the non-self-consistent
and self-consistent theories suggests that more accurate theoretical
descriptions are needed, in order to fully account for the current
spectroscopic observations on Fermi polarons.
\end{abstract}
\maketitle

\section{Introduction}

Fermi polaron, an impurity interacting with a non-interacting Fermi
sea of fermions, is probably the oldest and simplest quantum many-body
system that plays a significant role in our understanding of many-particle
physics \cite{Landau1933,Alexandrov2010}. Recent rapid experimental
advances in cold-atom research have brought renewed interest in Fermi
polaron problem \cite{Chevy2006,Schirotzek2009,Massignan2014,Schmidt2018,Wang2023AB,Tajima2024AB},
due to the unprecedented tunability in interparticle interaction,
purity and dimensionality \cite{Bloch2008,Chin2010}. As a consequence,
the physics of Fermi polarons can now be experimentally explored in
a quantitative manner with atomic Fermi-Fermi mixtures or Bose-Fermi
mixtures near Feshbach resonances, in which in the dilute limit minority
fermionic or bosonic atoms act as independent, uncorrelated impurities
\cite{Schirotzek2009,Massignan2014}.

Spectroscopic measurements, such as the radio-frequency (rf) spectroscopy
\cite{Schirotzek2009,Zhang2012,Kohstall2012,Koschorreck2012,Scazza2017,Zan2019},
Ramsey interferometry \cite{Scazza2017,Cetina2016}, and most recently
Raman spectroscopy \cite{Ness2020}, provide useful tools to reveal
a number of intriguing features of Fermi polarons. To date, polaron
energy has been measured from various spectroscopies with an excellent
accuracy and has been well explained by existing theories based on
variational Chevy ansatz \cite{Chevy2006,Cui2010,Parish2013,Liu2019,Liu2020,Hu2023AB},
diagrammatic many-body $T$-matrix approximations \cite{Combescot2007,Hu2018,Tajima2018,Wang2019,Mulkerin2019,Tajima2019,Hu2022,Hu2022b,Hu2022c,Hu2023},
and quantum Monte Carlo simulations \cite{Prokofev2008,Vlietinck2013,Kroiss2015}.
The dynamical properties concerning the measured spectroscopy lineshape,
however, are less understood. In particular, the spectral function
of Fermi polarons, which is the fundamental quantity that determines
the rf spectroscopy and Raman spectroscopy, is notoriously difficult
to accurately predict.

Pioneering quantum Monte Carlo simulations of the spectral function
have been attempted by Goulko and her collaborators \cite{Goulko2016}.
However, the accuracy need to be improved. Exact numerical calculations
are available in the heavy polaron limit of infinitely large impurity
mass \cite{Knap2012,Wang2022PRL,Wang2022PRA}. Yet, the experimental
realization of heavy Fermi polarons are to be demonstrated. Current
knowledge of the polaron spectral function largely relies on a non-self-consistent
many-body $T$-matrix theory \cite{Tajima2018,Mulkerin2019,Tajima2019,Hu2022,Hu2022b,Hu2022c,Hu2023}
or its equivalent form of variational Chevy ansatz \cite{Liu2019,Liu2020},
both at zero temperature and finite temperature. In the non-self-consistent
$T$-matrix theory, the successive scatterings between the impurity
and Fermi sea are taken into account in the form of ladder diagrams,
whose contributions can be diagrammatically calculated by using the
bare, non-interacting impurity Green function \cite{Combescot2007,Hu2022}.

The purpose of this work is to calculate the spectral function of
Fermi polarons based on a self-consistent many-body $T$-matrix theory,
in which the contributions from ladder diagrams are self-consistently
calculated by using a dressed, interacting impurity Green function.
A similar theoretical investigation was presented earlier by Tajima
and his co-workers \cite{Tajima2019}, where numerical calculations
were carried out with imaginary-time Green functions to avoid numerical
instability. We improve their interesting work by using real-time
Green functions. This may remove potential errors due to the uncontrollable
numerical analytic continuation applied to convert imaginary frequency
to real frequency, which is known to be ill-defined \cite{Goulko2016}.

We observe that the non-self-consistent and self-consistent theories
lead to quantitatively different predictions for the spectral function
of Fermi polarons. As a result, the predicted rf spectrum and Raman
spectrum also differ quantitatively. In comparison with the most recent
spectroscopic measurements \cite{Zan2019,Ness2020}, both predictions
can not explain the experimental data in a satisfactory way, although
the self-consistent results seem to provide a slightly better agreement.
The discrepancy between theories and experiments emphasizes the importance
of developing a more accurate theoretical framework for Fermi polarons.

It should be noted that, in perturbative diagrammatic theories, the
advantage of considering self-consistency in Feynman diagrams is not
taken for granted, particularly in the strongly interacting regime
that we are exploring \cite{Haussmann1994,Liu2005,Hu2008}. For example,
for a strong-interacting balanced spin-1/2 Fermi gas with equal spin
population, both the non-self-consistent and self-consistent many-body
$T$-matrix theories have been used to calculate the spectral function
of a unitary Fermi gas with infinitely large scattering length at
Feshbach resonance \cite{Chen2004,Tsuchiya2009,Haussmann2009,Palestini2012}.
However, the accuracy of both calculations receives unsettled debates,
since they lead to entirely different predictions on the existence
of pair-fluctuation-induced pesudogap \cite{Mueller2017,Li2024}.
A possible source for this qualitative discrepancy may arise from
the errors in numerical analytic continuation adopted in the self-consistent
$T$-matrix calculations \cite{Haussmann2009}. It would be interesting
to remove such avoidable errors in the self-consistent theory, by
extending our work to directly calculate the spectral function of
the unitary Fermi gas in real frequency.

The rest of the paper is organized as follows. In the next section
(Sec. II), we outline the model Hamiltonian for Fermi polarons and
briefly summarize the self-consistent many-body $T$-matrix approach.
We emphasize how to realize the numerical procedure for self-consistency
of the impurity Green function, with real frequency. In Sec. III,
we discuss the vertex function, the impurity self-energy and the polaron
spectral function and show the changes in these quantities because
of our self-consistent treatment. We also present the temperature
dependence of the polaron energy and decay rate. In Sec. IV, we calculate
the rf spectrum and Raman spectrum. We compare the theoretical results,
predicted by both non-self-consistent and self-consistent many-body
$T$-matrix theories, with the experimental data. The conclusions
and outlooks follow in Sec. V.

\section{Model Hamiltonian and self-consistent many-body $T$-matrix approach}

As in the experiments \cite{Schirotzek2009,Scazza2017,Zan2019,Ness2020},
we consider a highly imbalanced spin-1/2 Fermi gas of ultracold atoms
with equal mass $m$ near an $s$-wave Feshbach resonance, distributed
uniformly in volume $V$ in three dimensions. In the limit of vanishing
density of minority atoms, we treat them as uncorrelated impurities,
interacting with a non-interacting Fermi sea of majority atoms via
a contact interaction potential $g\delta(\mathbf{r}-\mathbf{r}')$.
Here, $g$ is the bare interaction strength that has to be replaced
by the $s$-wave scattering length $a$ using the standard relation,
\begin{equation}
\frac{1}{g}=\frac{m}{4\pi\hbar^{2}a}-\frac{1}{V}\sum_{\mathbf{k}}\frac{m}{\hbar^{2}\mathbf{k}^{2}},
\end{equation}
so the ultraviolet divergence inherent in the contact potential can
be effectively regularized. The system under consideration is well-described
by a single-channel model Hamiltonian,

\begin{equation}
\mathcal{H}=\sum_{\mathbf{k}}\epsilon_{\mathbf{k}}c_{\mathbf{k}}^{\dagger}c_{\mathbf{k}}+\sum_{\mathbf{k}}\epsilon_{\mathbf{k}}d_{\mathbf{k}}^{\dagger}d_{\mathbf{k}}+g\sum_{\mathbf{qkk'}}c_{\mathbf{k}}^{\dagger}d_{\mathbf{q}-\mathbf{k}}^{\dagger}d_{\mathbf{q-k'}}c_{\mathbf{k'}},
\end{equation}
where $c_{\mathbf{k}}^{\dagger}$ ($c_{\mathbf{k}}$) and $d_{\mathbf{k}}^{\dagger}$
($d_{\mathbf{k}}$)are the creation (annihilation) field operators
for fermionic atoms and the impurity, respectively. For clarity, we
have suppressed the volume $V$ in the model Hamiltonian, so the integration
over the momentum $\sum_{\mathbf{k}}$ in the following should be
always understood as $(1/V)\sum_{\mathbf{k}}=\int d\mathbf{k}/(2\pi)^{3}$.
The first two terms in the Hamiltonian describe the kinetic, non-interacting
part with the dispersion relation $\epsilon_{\mathbf{k}}=\hbar^{2}\mathbf{k}^{2}/(2m)$,
and the last term describe the interaction between the impurity and
Fermi sea. The chemical potentials are not specified in the Hamiltonian,
but it should be understood that the number $n$ of fermions in the
Fermi sea is tuned by a chemical potential $\mu$, i.e., we will modify
the single-particle dispersion relation to $\xi_{\mathbf{k}}=\epsilon_{\mathbf{k}}-\mu=\hbar^{2}\mathbf{k}^{2}/(2m)-\mu$.
Moreover, for a single impurity, it is not necessary to explicitly
introduce an impurity chemical potential \cite{Combescot2007,Hu2022}.
Throughout the work, we will take the Fermi wavevector $k_{F}=(6\pi^{2}n)^{1/3}$
and the Fermi energy $\varepsilon_{F}=\hbar^{2}k_{F}^{2}/(2m)$ as
the units of the wavevector $k$ (or $q$) and of the energy (or frequency),
respectively.

\subsection{Many-body $T$-matrix theories}

We solve the model Hamiltonian by using the many-body $T$-matrix
theories, which are well-documented in the literature \cite{Combescot2007,Hu2018,Tajima2019,Hu2022}.
Here, we only summarize the key equations, which are relevant to address
the self-consistency of the impurity Green function that we wish to
focus in this work. 

In the many-body $T$-matrix approximation, one keeps track on ladder
diagrams, which represent the successive forward scatterings between
the impurity and fermions in the particle-particle channel. At a nonzero
temperature $T$, the contributions of ladder diagrams are represented
by the inverse two-particle vertex function,
\begin{equation}
\Gamma^{-1}\left(\mathbf{q},\omega\right)=\frac{1}{g}-\sum_{\mathbf{k}}f\left(-\xi_{\mathbf{q-k}}\right)G\left(\mathbf{k},\omega-\xi_{\mathbf{q-k}}\right),\label{eq:vertexfunction}
\end{equation}
where $f(x)\equiv1/(e^{\beta x}+1)$ with $\beta\equiv1/(k_{B}T)$
is the Fermi-Dirac distribution function, and $G(\mathbf{k},\omega)$
is the retarded impurity Green function at momentum $\mathbf{k}$
with real frequency $\omega$ at finite temperature. In our self-consistent
treatment, this impurity Green function itself already includes the
interaction effect. In other words, it is a \emph{dressed} Green function
given by the Dyson equation, 
\begin{equation}
G\left(\mathbf{k},\omega\right)=\frac{1}{\omega-\epsilon_{\mathbf{k}}-\Sigma\left(\mathbf{k},\omega\right)},\label{eq:impurityGF}
\end{equation}
where the retarded impurity self-energy $\Sigma(\mathbf{k},\omega)$
is related to the vertex function $\Gamma(\mathbf{q},\omega)$,
\begin{equation}
\Sigma\left(\mathbf{k},\omega\right)=\sum_{\mathbf{q}}f\left(\xi_{\mathbf{q}-\mathbf{k}}\right)\Gamma\left(\mathbf{q},\omega+\xi_{\mathbf{q}-\mathbf{k}}\right).\label{eq:selfenergy}
\end{equation}
Eqs. (\ref{eq:vertexfunction}), (\ref{eq:impurityGF}) and (\ref{eq:selfenergy})
provide a set of coupled equations in the real-frequency domain, where
the dressed impurity Green function $G(\mathbf{k},\omega)$ needs
to be self-consistently determined.

In the non-self-consistent $T$-matrix theory, such a self-consistency
is not required. In Eq. (\ref{eq:vertexfunction}), we directly use
the non-interacting Green function $G_{0}(\mathbf{k},\omega-\xi_{\mathbf{q-k}})=1/(\omega-\xi_{\mathbf{q-k}}-\epsilon_{\mathbf{k}})$
to replace the dressed Green function $G(\mathbf{k},\omega-\xi_{\mathbf{q-k}})$,
yielding the expression \cite{Combescot2007},
\begin{equation}
\Gamma_{0}^{-1}\left(\mathbf{q},\omega\right)=\frac{m}{4\pi\hbar^{2}a}-\sum_{\mathbf{k}}\left[\frac{1-f\left(\xi_{\mathbf{q-k}}\right)}{\omega-\xi_{\mathbf{q-k}}-\epsilon_{\mathbf{k}}}+\frac{m}{\hbar^{2}\mathbf{k}^{2}}\right],\label{eq:vertexfunction0}
\end{equation}
where we have rewritten the bare interaction strength $g$ in terms
of the physical $s$-wave scattering length $a$. In turn, we substitute
the leading-order approximated vertex function $\Gamma_{0}(\mathbf{q},\omega)$
into Eq. (\ref{eq:selfenergy}), to determine the impurity self-energy
$\Sigma_{0}(\mathbf{k},\omega)$ at the \emph{first} iteration. In
the non-self-consistent treatment, we simply assume that the resulting
self-energy $\Sigma_{0}(\mathbf{k},\omega)$ might already be useful
enough and would lead to reasonably accurate impurity Green function
$G(\mathbf{k},\omega)$, when it is used in the Dyson equation Eq.
(\ref{eq:impurityGF}). 

In contrast, in our fully self-consistent treatment, we need to use
$\Sigma_{0}(\mathbf{k},\omega)$ to obtain an improved impurity Green
function $G_{1}(\mathbf{k},\omega-\xi_{\mathbf{q-k}})=1/[\omega-\xi_{\mathbf{q-k}}-\epsilon_{\mathbf{k}}-\Sigma_{0}(\mathbf{k},\omega-\xi_{\mathbf{q-k}})]$,
and then repeat the above-mentioned procedure to iteratively update
the impurity Green function, until it converges. The numerical workload
of self-consistent calculations is therefore much heavier.

\subsection{Numerical calculations}

To reduce the workload, it is worth noting that the key difficulty
of numerical calculations comes from the integration over the momentum
$\mathbf{k}$ in Eq. (\ref{eq:vertexfunction}), due to the poles
of the impurity Green function that makes the integrand very singular.
This singularity actually already appears in the non-self-consistent
calculations. As can be readily seen from Eq. (\ref{eq:vertexfunction0}),
the integrand on the right-hand side badly diverges at some momenta
$\mathbf{k}$, once the frequency $\omega$ is in the two-particle
continuum and satisfies $\omega=\xi_{\mathbf{q-k}}+\epsilon_{\mathbf{k}}$.
Fortunately, since we can precisely locate the pole position of the
non-interacting impurity Green function, the vertex function $\Gamma_{0}(\mathbf{q},\omega)$
can be efficiently calculated, as outlined in detail in the previous
work \cite{Hu2022}. As we anticipate that $\Gamma_{0}(\mathbf{q},\omega)$
makes the dominant contribution to the full vertex function $\Gamma(\mathbf{q},\omega)$,
the difference\begin{widetext}
\begin{equation}
\delta\chi\left(\mathbf{q},\omega\right)\equiv\Gamma^{-1}\left(\mathbf{q},\omega\right)-\Gamma_{0}^{-1}\left(\mathbf{q},\omega\right)=-\sum_{\mathbf{k}}\frac{f\left(-\xi_{\mathbf{q-k}}\right)\Sigma\left(\mathbf{k},\omega-\xi_{\mathbf{q-k}}\right)}{\left(\omega-\xi_{\mathbf{q-k}}-\epsilon_{\mathbf{k}}\right)\left[\omega-\xi_{\mathbf{q-k}}-\epsilon_{\mathbf{k}}-\Sigma\left(\mathbf{k},\omega-\xi_{\mathbf{q-k}}\right)\right]}\label{eq:dkappa}
\end{equation}
\end{widetext}would be small and therefore does not require high-precision
calculation. We may then artificially introduce a small imaginary
part $\eta$ to the real frequency $\omega$ to remove the singularity
in the integrand of Eq. (\ref{eq:dkappa}). In practice, we find that
the approximated expression,
\begin{equation}
\delta\chi\left(\mathbf{q},\omega\right)\simeq2\delta\chi\left(\mathbf{q},\omega+i\eta\right)-\delta\chi\left(\mathbf{q},\omega+2i\eta\right),\label{eq:dkappaAppr}
\end{equation}
works extremely well, with a small $\eta=0.2\varepsilon_{F}$. The
choice of this value for $\eta$ has been carefully examined in Appendix
A. We have confirmed that the converged results of the impurity Green
function and spectral function do not depend on $\eta$.

The whole procedure of numerical iterations is then simple to carry
out. We start from the non-self-consistent result of the self-energy
$\Sigma_{0}(\mathbf{k},\omega)$ and calculate the difference $\delta\chi(\mathbf{q},\omega)$
using Eq. (\ref{eq:dkappa}) and Eq. (\ref{eq:dkappaAppr}). We then
update the vertex function,
\begin{equation}
\Gamma\left(\mathbf{q},\omega\right)=\frac{1}{\Gamma_{0}^{-1}\left(\mathbf{q},\omega\right)+\delta\chi\left(\mathbf{q},\omega\right)},
\end{equation}
and use it to obtain a new self-energy $\Sigma(\mathbf{k},\omega)$
with Eq. (\ref{eq:selfenergy}). The iteration is repeated until the
change $\delta\Sigma(\mathbf{k},\omega)$ in the self-energy becomes
negligible. Typically, the convergence can be quickly reached in just
a few iterations. During the iteration procedure, the self-energy
$\Sigma(\mathbf{k},\omega)$ will be stored in the form of a two-dimensional
array. The numbers of grid points for the momentum $k=\left|\mathbf{k}\right|$
and the frequency $\omega$ are about $200$ and $500$, respectively.
The dense grid points are distributed in a non-equidistant way, so
both large-momentum and large-frequency behaviors of the self-energy
can be well sampled. We can then use a cubic spline interpolation
to accurately extract a self-energy $\Sigma(\mathbf{k},\omega)$ at
arbitrary momentum $k$ and frequency $\omega$.

It is also worth noting that, for a large and positive dimensionless
interaction parameter $1/(k_{F}a)$, the vertex function $\Gamma(\mathbf{q},\omega)$
may develop a pole, which signals the existence of a well-defined
molecule state \cite{Combescot2007}. In that case, specific attention
should be paid to handle the singularity in Eq. (\ref{eq:selfenergy}).
However, in this work, we always focus on the polaron regime, where
the vertex function $\Gamma(\mathbf{q},\omega)$ is consistently well
behaved. Although the initial vertex function $\Gamma_{0}(\mathbf{q},\omega)$
may suffer from a singularity at large $1/(k_{F}a)$ in the polaron
regime, the singularity will be quickly removed by the self-consistency
iteration.

\begin{figure}
\begin{centering}
\includegraphics[width=0.5\textwidth]{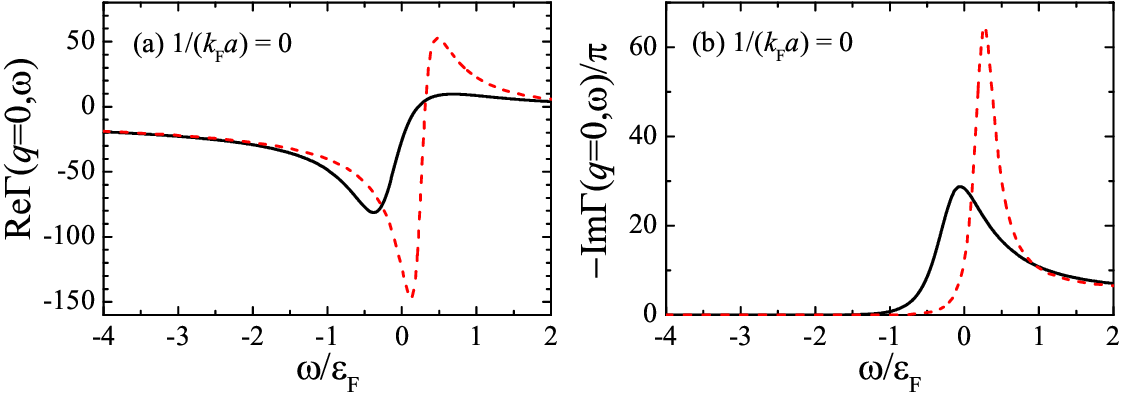}
\par\end{centering}
\begin{centering}
\includegraphics[width=0.5\textwidth]{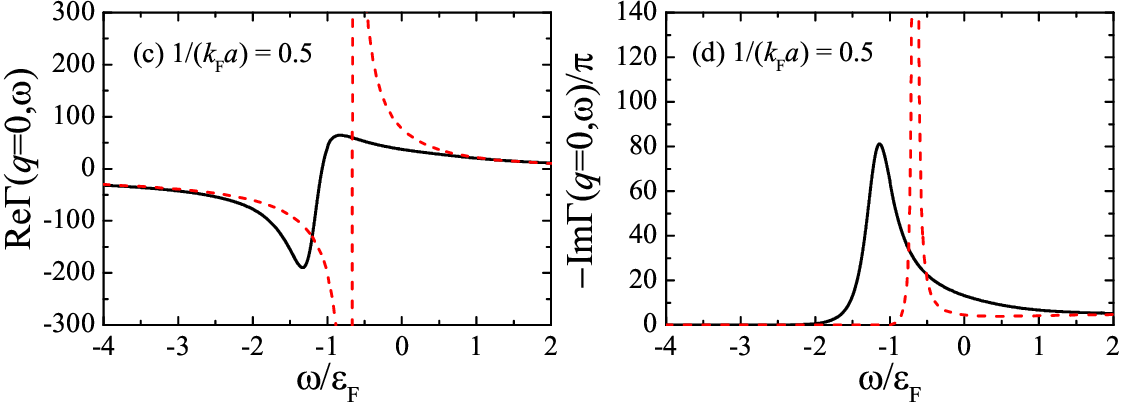}
\par\end{centering}
\caption{\label{fig:fig1_vertexfunction} The real part (a, c) and imaginary
part (b, d) of the zero-momentum vertex function $\Gamma(q=0,\omega)$,
in arbitrary units, at the temperature $T=0.2T_{F}$. The two upper
panels are the results in the unitary limit (i.e., $1/(k_{F}a)=0$),
while the two bottom panels present the results at the molecular side
of the Feshbach resonance with $1/(k_{F}a)=0.5$. The black solid
lines and the red dashed lines correspond to the predictions from
the self-consistent and non-self-consistent many-body $T$-matrix
theories, respectively.}
\end{figure}

\section{Polaron spectral function}

In Fig. \ref{fig:fig1_vertexfunction}, we report the zero-momentum
vertex function at two interaction parameters and at a low temperature
$T=0.2T_{F}$, predicted by either the self-consistent (black solid
lines) or non-self-consistent (red dashed lines) many-body $T$-matrix
theories. In the upper panel of the figure, we take the unitary limit
$1/(k_{F}a)=0$, where a two-body bound state starts to emerge. In
the lower panel, we consider the BEC (Bose-Einstein condensate) side
or the molecule side of the Feshbach resonance with $1/(k_{F}a)=0.5$,
where a two-body bound state exists, with binding energy $E_{B}=2\varepsilon_{F}/(k_{F}a)^{2}=0.5\varepsilon_{F}$.
For both interaction parameters, we find that the self-consistency
treatment strongly modifies the results of the vertex function.

Let us focus on the imaginary part of the vertex function shown in
Fig. \ref{fig:fig1_vertexfunction}(b) and Fig. \ref{fig:fig1_vertexfunction}(d).
Physically, the vertex function describes a molecule state in the
presence of the many-body environment of a Fermi sea \cite{Prokofev2008}.
Its imaginary part can therefore be used to define a molecule spectral
function,
\begin{equation}
A_{\textrm{mol}}\left(\mathbf{q},\omega\right)\propto-\frac{1}{\pi}\textrm{Im}\Gamma\left(\mathbf{q},\omega\right),
\end{equation}
which is precisely the quantity plotted in the figure. We always find
a peak in $A_{\textrm{mol}}(\mathbf{q},\omega)$, although there is
no well-defined two-body bound state in the unitary limit. This is
understandable, since the presence of a Fermi sea is known to be favorable
for stabilizing a many-body (Cooper) pair \cite{Cooper1956}. It is
easy to see that the non-self-consistent $T$-matrix theory predicts
a much sharper peak in the molecule spectral function than the self-consistent
$T$-matrix theory. Moreover, with the self-consistency in the impurity
Green function, the molecule peak shifts to the low energy side, by
an amount about $0.5\varepsilon_{F}$.

Although for the impurity Green function the advantage of taking the
self-consistency is not granted, for the vertex function (or boldly
the molecule Green function), there is no doubt that the self-consistency
treatment will improve its accuracy. The lower molecule peak or smaller
molecule energy, predicted by the self-consistent $T$-matrix theory,
therefore implies that the critical interaction strength for the polaron-molecule
transition \cite{Punk2009} can be smaller than what predicted by
the non-self-consistent $T$-matrix theory. This observation is consistent
with the previous $T$-matrix studies for polaron energy at zero temperature
\cite{Hu2018}.

\begin{figure}
\begin{centering}
\includegraphics[width=0.5\textwidth]{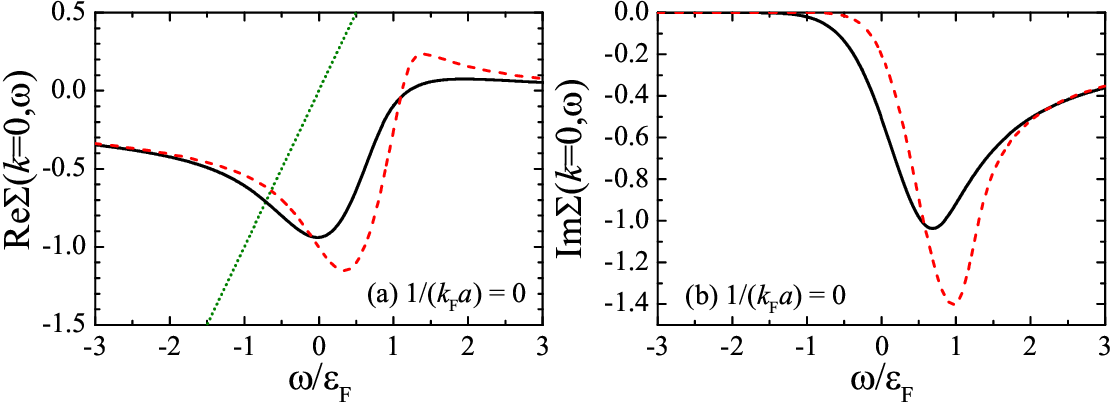}
\par\end{centering}
\begin{centering}
\includegraphics[width=0.5\textwidth]{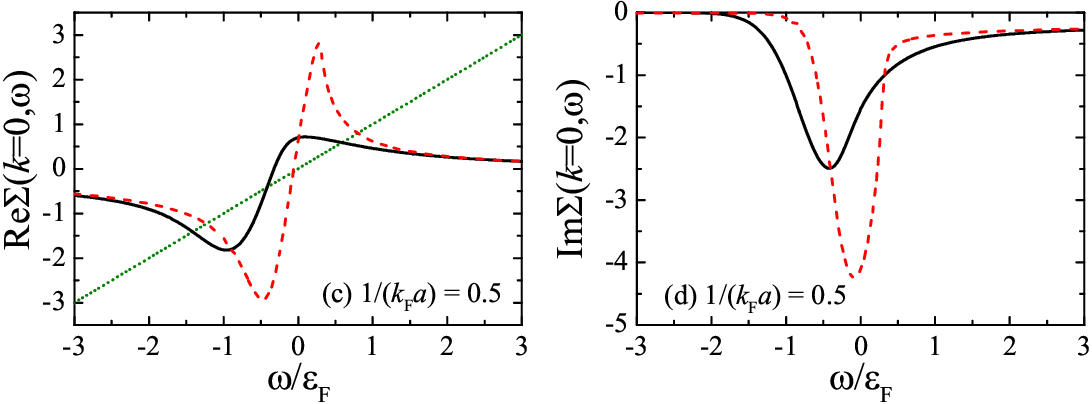}
\par\end{centering}
\caption{\label{fig:fig2_selfenergy} The real part (a, c) and imaginary part
(b, d) of the zero-momentum self-energy $\Sigma(k=0,\omega)$, in
units of $\varepsilon_{F}$, at the temperature $T=0.2T_{F}$ and
at the two interaction strengths as indicated. The black solid lines
and the red dashed lines correspond to the predictions from the self-consistent
and non-self-consistent many-body $T$-matrix theories, respectively.
In (a) and (c), the cross points of $\textrm{Re}\Sigma(k=0,\omega)$
and the green dotted curves (i.e., $y=\omega$) determine the polaron
energies (of different polaron branches).}
\end{figure}

In Fig. \ref{fig:fig2_selfenergy}, we present the real part and imaginary
part of the impurity self-energy at zero momentum, and at the same
parameters as in Fig. \ref{fig:fig1_vertexfunction}. Once again,
we find significant changes due to the self-consistency treatment.
At $k=0$, the pole of the impurity Green function in Eq. (\ref{eq:impurityGF})
occurs at $\omega=\textrm{Re}\Sigma(k=0,\omega)$, if we neglect the
(possibly large) imaginary part $\textrm{Im}\Sigma(k=0,\omega)$.
Therefore, in Fig. \ref{fig:fig2_selfenergy}(a) and Fig. \ref{fig:fig2_selfenergy}(c),
we also show the curve $y=\omega$ in a green dotted line. The cross
point between the green dotted line and the curve $\textrm{Re}\Sigma(k=0,\omega)$
determines the polaron energy $\mathcal{E}_{P}$ at the pole of the
impurity Green function. 

\begin{figure}
\begin{centering}
\includegraphics[width=0.5\textwidth]{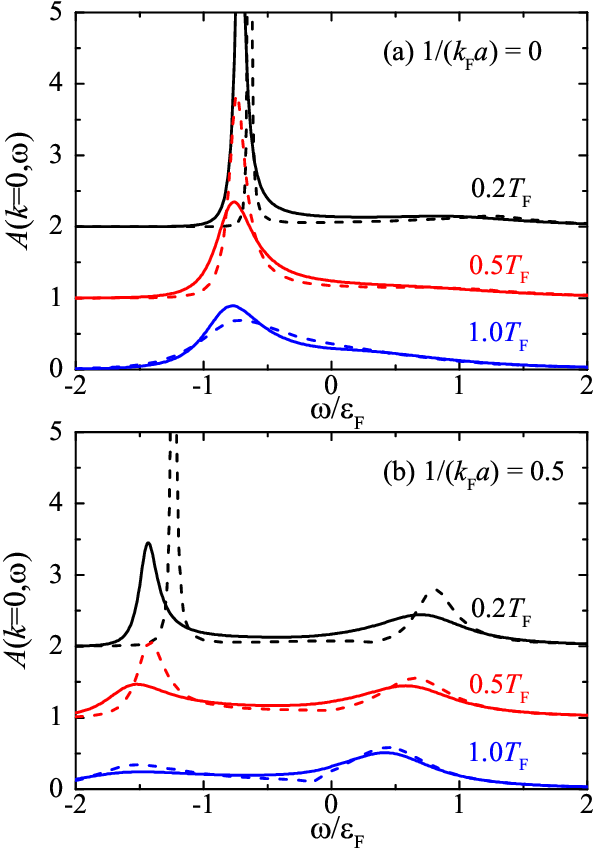}
\par\end{centering}
\caption{\label{fig:fig3_akw} Zero-momentum spectral function of the impurity
(in units of $\varepsilon_{F}^{-1}$) in the unitary limit (a) and
on the molecular side of the Feshbach resonance (b), at three different
temperatures as indicated. For clarity, the results at $T=0.2T_{F}$
and $T=0.5T_{F}$ have been vertically up-shifted. The solid lines
(dashed lines) show the predictions of the self-consistent (non-self-consistent)
many-body $T$-matrix theory.}
\end{figure}

On the negative frequency side, we always find a cross point, which
gives the energy of attractive Fermi polaron. The self-consistent
many-body $T$-matrix theory predicts a lower polaron energy than
its non-self-consistent counterpart, with an energy shift about $0.1\sim0.2\varepsilon_{F}$
that is smaller than the shift in the molecule energy that we observe
in the molecule spectral function. On the other hand, on the positive
frequency side, we can only find the cross point at the positive interaction
parameter $1/(k_{F}a)>0$ in Fig. \ref{fig:fig2_selfenergy}(c), which
determines the energy of repulsive Fermi polaron. As in the case of
the attractive polaron, we observe that the self-consistency still
leads to a small red-shift in the repulsive polaron energy. Interestingly,
at $1/(k_{F}a)=0.5$ there is an additional cross point located close
to the zero frequency $\omega=0$. However, this cross point can hardly
be viewed as the pole of the impurity Green function, since the imaginary
part of the impurity self-energy becomes too large near $\omega=0$.

For the imaginary part of the impurity self-energy shown in \ref{fig:fig2_selfenergy}(b)
and Fig. \ref{fig:fig2_selfenergy}(d), we find that the self-consistent
$T$-matrix theory consistently predicts a more negative imaginary
part than the non-self-consistent $T$-matrix theory, at the frequency
near the polaron energy. As we shall see, it will lead to the prediction
of a larger decay rate of polaron quasiparticles.

We now turn to discuss the impurity spectral function defined by 
\begin{equation}
A\left(\mathbf{k},\omega\right)=-\frac{1}{\pi}\textrm{Im}G\left(\mathbf{k},\omega\right).
\end{equation}
Near the polaron energy $\mathcal{E}_{P}$ at the pole of the impurity
Green function, we may Taylor-expand the self-energy at small momentum,
\begin{widetext}
\begin{equation}
\Sigma\left(\mathbf{k\rightarrow0},\omega\rightarrow\mathcal{E}_{P}\right)\simeq\mathcal{E}_{P}+\left.\frac{\partial\textrm{Re}\Sigma\left(\mathbf{k},\mathcal{E}_{P}\right)}{\partial\epsilon_{\mathbf{k}}}\right|_{\mathbf{k}=0}\epsilon_{\mathbf{k}}+\left.\frac{\partial\textrm{Re}\Sigma\left(0,\omega\right)}{\partial\omega}\right|_{\omega=\mathcal{E}_{P}}\left(\omega-\mathcal{E}_{P}\right)+i\textrm{Im}\Sigma\left(0,\mathcal{E}_{P}\right),
\end{equation}
\end{widetext}where we have used the condition $\mathcal{E}_{P}=\textrm{Re}\Sigma(0,\mathcal{E}_{P})$.
It is a convention to introduce the polaron residue 
\begin{equation}
\mathcal{Z}=\left[1-\left.\frac{\partial\textrm{Re}\Sigma(\mathbf{0},\omega)}{\partial\omega}\right|_{\omega=\mathcal{E}_{P}}\right]^{-1}
\end{equation}
and polaron decay rate 
\begin{equation}
\Gamma=-2\mathcal{Z}\textrm{Im}\Sigma\left(\mathbf{0},\mathcal{E}_{P}\right),
\end{equation}
with which we may explicitly rewrite the zero-momentum spectral function
$A(k=0,\omega)$ into the approximate Lorentzian form near the polaron
energy,
\begin{equation}
A\left(0,\omega\right)\simeq\mathcal{Z}\frac{\Gamma/\left(2\pi\right)}{\left(\omega-\mathcal{E}_{P}\right)^{2}+\Gamma^{2}/4}.
\end{equation}
Therefore, the residue $\mathcal{Z}$ measures the area under the
polaron peak and the decay rate $\Gamma$ determines the full width
at the half maximum (FWHM) of the peak. 

In Fig. \ref{fig:fig3_akw}, we report the zero-momentum impurity
spectral function at three typical temperatures in the unitary limit
with $1/(k_{F}a)=0$ (a) and on the BEC side of the Feshbach resonance
with $1/(k_{F}a)=0.5$ (b). In comparison with the non-self-consistent
$T$-matrix theoretical results (i.e., red dashed lines), it is readily
seen that the self-consistent $T$-matrix theory always predicts a
broader polaron peak, indicating a larger polaron decay rate. The
temperature evolution of the polaron spectral function provided by
the two $T$-matrix theories are qualitatively similar. However, there
are quantitative differences that we shall discuss in detail in the
following.

\begin{figure}
\begin{centering}
\includegraphics[width=0.5\textwidth]{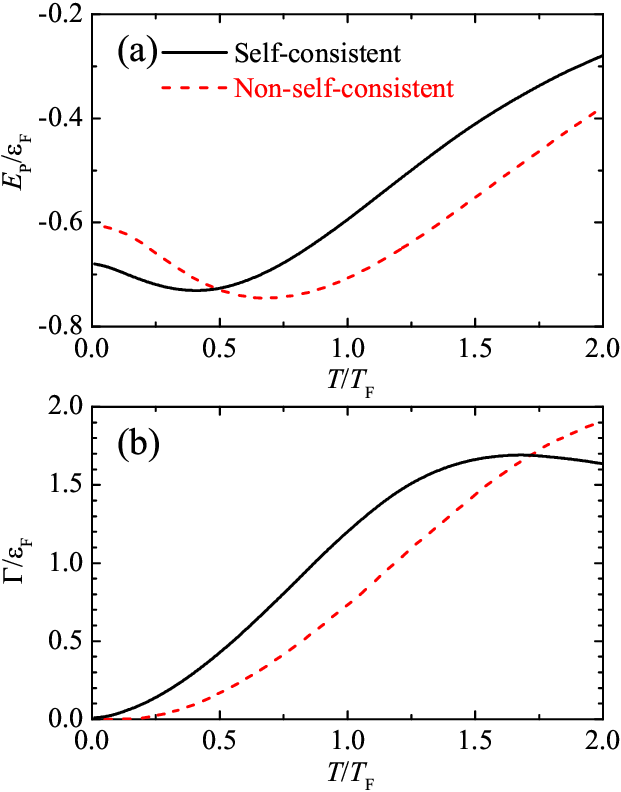}
\par\end{centering}
\caption{\label{fig:fig4_EpUnitary} Temperature dependence of the energy (a)
and decay rate (b) of the attractive polaron in the unitary limit.
The black solid lines and red dashed lines report the predictions
of the self-consistent and non-self-consistent many-body $T$-matrix
theories, respectively.}
\end{figure}

In the unitary limit (Fig. \ref{fig:fig3_akw}(a)), both $T$-matrix
theories show a non-monotonic temperature dependence of the (attractive)
polaron energy. This can be seen more clearly in Fig. \ref{fig:fig4_EpUnitary},
where we report the polaron energy and decay rate as a function of
the temperature. With increasing temperature, the polaron energy initially
decreases, reaches a global minimum at certain temperature and then
increases. The initial decrease in the polaron energy with temperature
might be understood from the Pauli exclusion principle. The thermal
blurring of the Fermi sea reduces the statistical exclusion and therefore
is favorable for the particle-hole excitations that are crucial for
the polaron formation \cite{Zan2019}. However, a large temperature
eventually reduces the effective interaction between the impurity
and Fermi sea, and increases the polaron energy. We find that the
two many-body $T$-matrix theories give different temperatures for
the minimum polaron energy, so the two curves of polaron energy cross
at around $T_{X}\sim0.5T_{F}$. Above $T_{X}$, the self-consistent
$T$-matrix theory predicts a larger polaron energy than the non-self-consistent
$T$-matrix theory, different from what we observe in Fig. \ref{fig:fig2_selfenergy}
for the impurity self-energy at $T=0.2T_{F}$.

\begin{figure}
\begin{centering}
\includegraphics[width=0.5\textwidth]{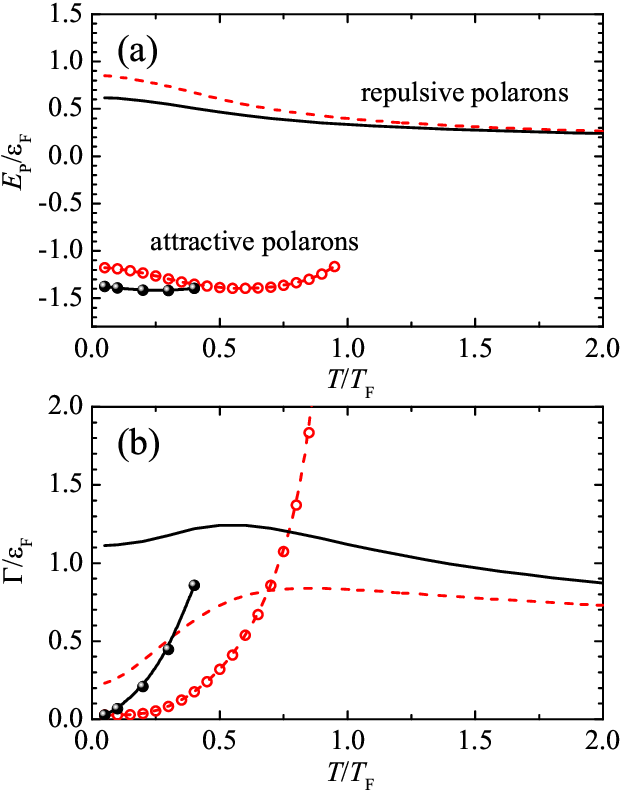}
\par\end{centering}
\caption{\label{fig:fig5_EpVx05p} Temperature dependence of the energy (a)
and decay rate of the attractive polaron and repulsive polaron at
the interaction strength $1/(k_{F}a)=0.5$. The black solid lines
and red dashed lines respectively report the predictions of the self-consistent
and non-self-consistent many-body $T$-matrix theories, for the repulsive
polaron. The black solid lines with solid circles and red dashed lines
with open circles correspond to the results of the attractive polaron.}
\end{figure}

On the BEC side of the Feshbach resonance (see Fig. \ref{fig:fig3_akw}(b)),
there are two branches of Fermi polarons. In Fig. \ref{fig:fig5_EpVx05p},
we show the temperature dependence of polaron energy and decay rate
for both attractive Fermi polaron and repulsive Fermi polaron. The
attractive polaron branch does not always exist. We can only find
the attractive polaron solution for $\mathcal{E}_{P}=\textrm{Re}\Sigma(0,\mathcal{E}_{P})$
at temperature $T<0.5T_{F}$ in the self-consistent many-body $T$-matrix
theory. The non-self-consistent theory seems to give a wider temperature
range for attractive polaron, i.e., $T<T_{F}$. However, its decay
rate increases too rapidly with temperature. As a result, it may hardly
be viewed as a well-defined quasiparticle once $T>0.7T_{F}$, where
the decay rate becomes larger than $\varepsilon_{F}$. 

In contrast, we can always find a solution of $\mathcal{E}_{P}=\textrm{Re}\Sigma(0,\mathcal{E}_{P})$
for the repulsive polaron branch. The repulsive polaron energy predicted
by the two theories does not differ too much. The difference is about
$0.3\varepsilon_{F}$ at most near zero temperature and it becomes
negligible above the Fermi degenerate temperature. Remarkably, the
decay rates of the repulsive polaron given by the two $T$-matrix
theories are very different. In particular, near zero temperature
the decay rate obtained from the self-consistent calculations is about
$1.1\varepsilon_{F}$, significantly larger than the non-self-consistent
result of about $0.2\varepsilon_{F}$. In comparison with the recent
measurement from LENS (the European Laboratory for Non-Linear Spectroscopy),
the decay rate of the repulsive polaron calculated from the non-self-consistent
$T$-matrix theory agrees better with the experimental data, which
are about tens of percent of the Fermi energy \cite{Scazza2017}.
We note finally that, the repulsive polaron energy from both $T$-matrix
theories decreases with increasing temperature. The temperature dependence
of the repulsive polaron decay rate from both theories is not monotonic;
but once $T>0.5T_{F}$, the decay rate becomes comparable to $\varepsilon_{F}$.

\section{Atomic spectroscopy}

In experiments, the polaron spectral function can be probed by using
rf spectroscopy or Raman spectroscopy. In those spectroscopic measurements,
the impurity is initially in the hyperfine state that interacts with
the Fermi sea. It is then transferred or ejected to a second, non-interacting
hyperfine state using either rf beams or Raman beams with energy $\omega$.
According to the linear response theory the ejection rate is proportional
to \cite{Zan2019,Ness2020,Hu2022,Hu2023},

\begin{equation}
I\left(\omega\right)=\frac{1}{V}\sum_{\mathbf{k}}A\left[\mathbf{k},\epsilon_{\mathbf{k}+\mathbf{Q}}-\omega\right]f\left(\epsilon_{\mathbf{k}+\mathbf{Q}}-\omega-\mu_{I}\right),\label{eq:ejectionSpectroscopy}
\end{equation}
where $\mathbf{Q}$ is the momentum of the light beams. In the case
of rf spectroscopy, the momentum is negligible, so we take $\mathbf{Q}=0$.
Realistically, one always uses a small impurity density $n_{\textrm{imp}}$
in the experiments \cite{Zan2019,Ness2020}, which can be theoretically
set by an impurity chemical potential $\mu_{I}$ through the number
equation,
\begin{equation}
n_{\textrm{imp}}=\frac{1}{V}\sum_{\mathbf{k}}\intop_{-\infty}^{+\infty}d\omega f\left(\omega-\mu_{I}\right)A\left(\mathbf{k},\omega\right).
\end{equation}
By integrating over the frequency in Eq. (\ref{eq:ejectionSpectroscopy}),
it is easy to see that the rf spectrum or Raman spectrum is normalized
to the impurity density, i.e., $\int d\omega I(\omega)=n_{\textrm{imp}}$.
In the following, we will alway plot a normalized spectrum by dividing
$I(\omega)$ by $n_{\textrm{imp}}$. It is also useful to note that,
to calculate the rf spectrum or Raman spectrum one needs to integrate
the spectral function over different momentum. As a result, a clear
interpretation of the spectroscopic measurement, in terms of zero-momentum
spectral function, may become difficult. In Appendix B, we briefly
discuss the spectral function of a unitary Fermi polaron at finite
momentum, predicted by the two $T$-matrix theories.

\begin{figure}
\begin{centering}
\includegraphics[width=0.5\textwidth]{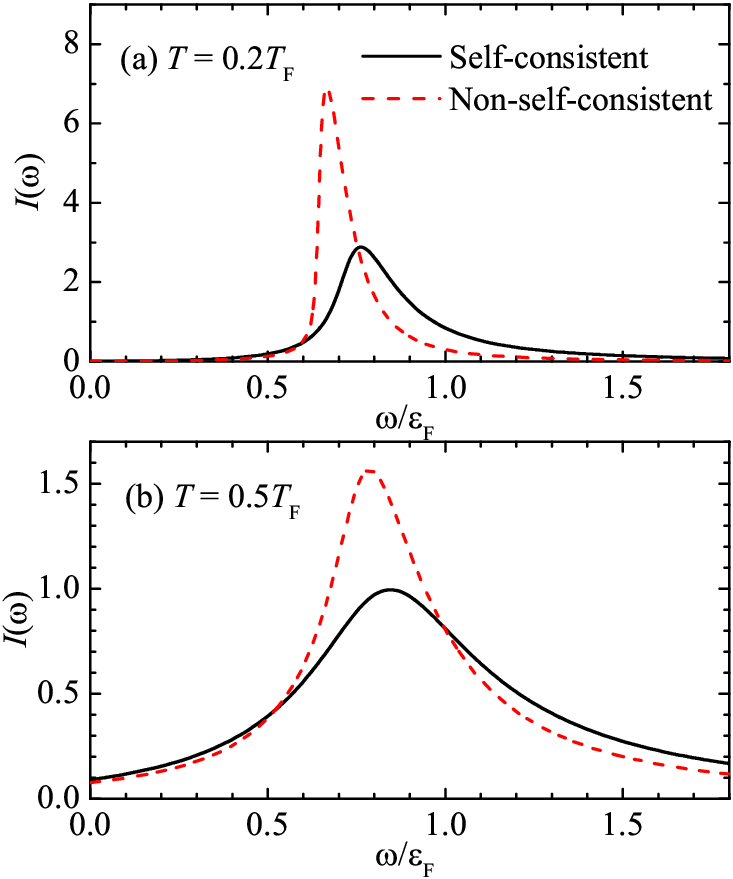}
\par\end{centering}
\caption{\label{fig:fig6_RFUnitary} The ejection rf spectra of a unitary Fermi
polaron at $T=0.2T_{F}$ (a) and $T=0.5T_{F}$ (b), in units of $\varepsilon_{F}^{-1}$.
The black solid lines and red dashed lines report the predictions
of the self-consistent and non-self-consistent many-body $T$-matrix
theories, respectively. Here, we use the impurity density $n_{\textrm{imp}}=0.1n$.
The spectrum is normalized, so $\int d\omega I(\omega)=1$. }
\end{figure}

\subsection{RF spectrum}

In Fig. \ref{fig:fig6_RFUnitary}, we show the rf spectra of a unitary
Fermi polaron at two temperatures, $T=0.2T_{F}$ (a) and $T=0.5T_{F}$
(b). In each spectrum, there is a peak associated with the attractive
polaron. The peak position locates at $\omega\simeq-\mathcal{E}_{P}$,
while the peak width might become broader than the decay rate of the
zero-momentum attractive polaron, due to the contribution from finite-momentum
Fermi polarons. We observe that the self-consistent many-body $T$-matrix
theory predicts a broader rf peak at higher energy than the non-self-consistent
theory. This is consistent with what we find in the spectral function. 

The changes due to the self-consistent treatment are quantitively
significant, as the height of the rf peak is much reduced. For example,
at the low temperature $T=0.2T_{F}$, the rf peak height can be reduced
by a factor of more than two. As temperature increases, the reduction
effect becomes weaker. At $T=0.5T_{F}$, we can only see a reduction
of about $35\%$.

\begin{figure}
\begin{centering}
\includegraphics[width=0.5\textwidth]{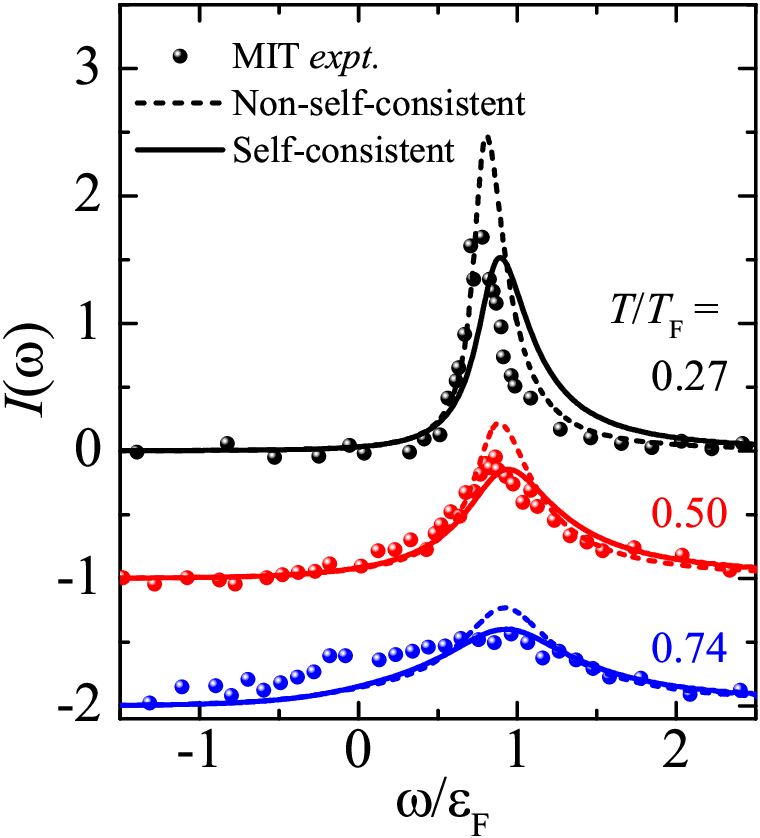}
\par\end{centering}
\caption{\label{fig:fig7_RFMITexpt1} The ejection rf spectra, predicted by
the self-consistent (solid lines) and non-self-consistent (dashed
lines) many-body $T$-matrix theories, are compared with the experimental
data from MIT (circles) \cite{Zan2019}. Here, we use the impurity
density $n_{\textrm{imp}}=0.1n$. We have applied a Lorentzian broadening
on all the theoretical curves to take into account a well-calibrated
experimental energy resolution $0.1\varepsilon_{F}$, and have also
shifted the curves to the right by an amount $0.09\varepsilon_{F}$,
in order to eliminate the residual final-state effect.}
\end{figure}

At this point, it is interesting to compare the theoretical predictions
of the two $T$-matrix theories with the latest rf measurement from
MIT (The Massachusetts Institute of Technology) at $T<T_{F}$, as
shown in Fig. \ref{fig:fig7_RFMITexpt1}. To simulate the realistic
experimental conditions, we have taken a convolution of the theoretical
rf spectrum with a Lorentzian lineshape, which accounts for the experimental
energy resolution of $0.1\varepsilon_{F}$. We have also horizontally
shifted the spectrum by an amount $0.09\varepsilon_{F}$, to compensate
the final-state effect arising from the residual interaction for the
impurity in the second hyperfine state after transfer. Otherwise,
there are no free adjustable parameters used in the comparison. 

It is readily seen that, overall the predictions from the self-consistent
many-body $T$-matrix theory fit better with the experimental data.
The non-self-consistent theory always predicts a higher peak height
than the experimental observation. It is worth noting that, as temperature
increases, a pronounced peak starts to emerge at about $\omega\sim0$
in the measured rf spectrum. However, both $T$-matrix theories fail
to produce such an important experimental feature. As a result, at
$T=0.74T_{F}$ we find an apparent discrepancy between theory and
experiment near the zero frequency (see the bottom curves and data
in Fig. \ref{fig:fig7_RFMITexpt1} at $\omega\sim0$).

\begin{figure}
\begin{centering}
\includegraphics[width=0.5\textwidth]{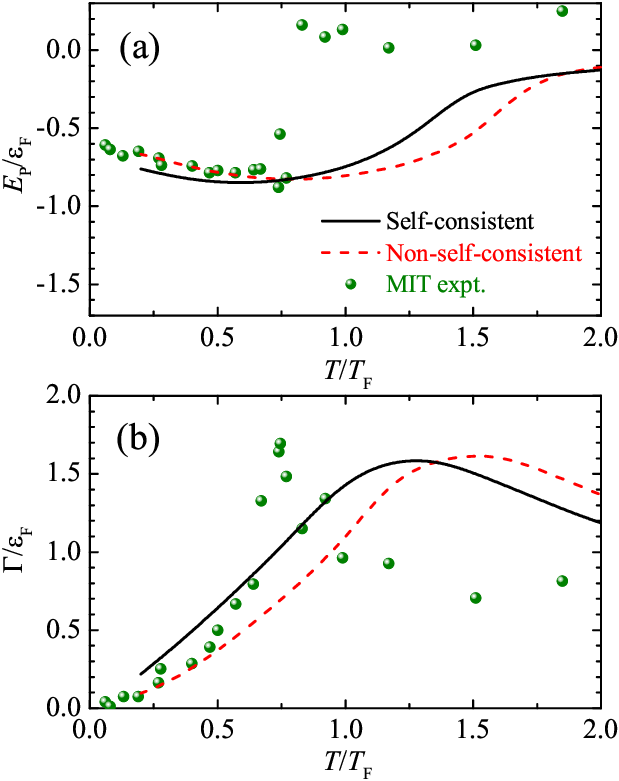}
\par\end{centering}
\caption{\label{fig:fig8_RFMITexpt2} The peak position and the FWHM width
of the ejection rf spectra, predicted by the self-consistent (black
solid lines) and non-self-consistent (red dashed lines) many-body
$T$-matrix theories, are compared with the experimental data from
MIT (green circles) \cite{Zan2019}. Here, we use the impurity density
$n_{\textrm{imp}}=0.1n$ in the calculations of the ejection rf spectra.
In the experimental data, we have subtracted the energy resolution
$0.1\varepsilon_{F}$ in the FWHM width $\Gamma$, and have compensated
the final-state energy shift $0.09\varepsilon_{F}$ in the extracted
peak position $-\mathcal{E}_{P}$ \cite{Zan2019}.}
\end{figure}

This discrepancy becomes particularly evident, when we compare the
results on the peak position (which is interpreted as $-\mathcal{E}_{P}$)
and the FWHM width (i.e., the decay rate $\Gamma$ of the polaron
in our interpretation), which are extracted from the simulated theoretical
curves and the measured rf spectra. These results are shown in Fig.
\ref{fig:fig8_RFMITexpt2}(a) and Fig. \ref{fig:fig8_RFMITexpt2}(b),
respectively. We see a clear jump in the measured peak position at
temperature $T\sim0.8T_{F}$. Above this temperature, the peak position
seems to be pinned near the zero frequency. In contrast, the peak
position predicted by both $T$-matrix theories gradually decreases
towards $\omega=0$. In line with the sudden jump in the peak position,
the measured FWHM width reaches maximum at $T\sim0.8T_{F}$. After
this temperature, the width quickly decreases. This observation also
cannot be understood by both $T$-matrix theories. As temperature
increases, there are maximum widths, occurring at $T\sim1.2T_{F}$
and $T\sim1.5T_{F}$ from the self-consistent and non-self-consistent
calculations, respectively. Moreover, the predicted temperature dependence
of the FWHM width appear to be much smoother than what observed in
the measured spectra.

\subsection{Raman spectrum}

\begin{figure}
\begin{centering}
\includegraphics[width=0.5\textwidth]{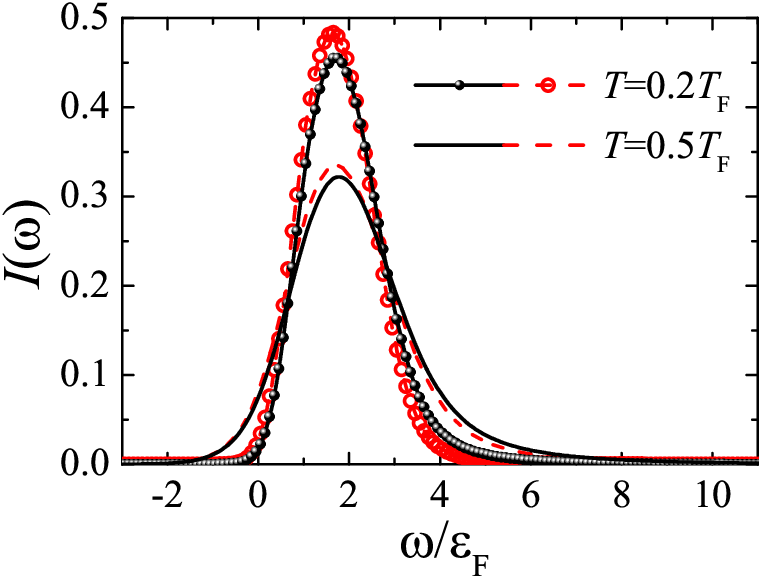}
\par\end{centering}
\caption{\label{fig:fig9_RamanUnitary} The ejection Raman spectra of a unitary
Fermi polaron at $T=0.2T_{F}$ (the lines with symbols) and $T=0.5T_{F}$
(the lines), in units of $\varepsilon_{F}^{-1}$. The black solid
lines and red dashed lines report the predictions of the self-consistent
and non-self-consistent many-body $T$-matrix theories, respectively.
Here, we use the transferred momentum $q=k_{F}$ and the impurity
density $n_{\textrm{imp}}=0.15n$.}
\end{figure}

\begin{figure}
\begin{centering}
\includegraphics[width=0.5\textwidth]{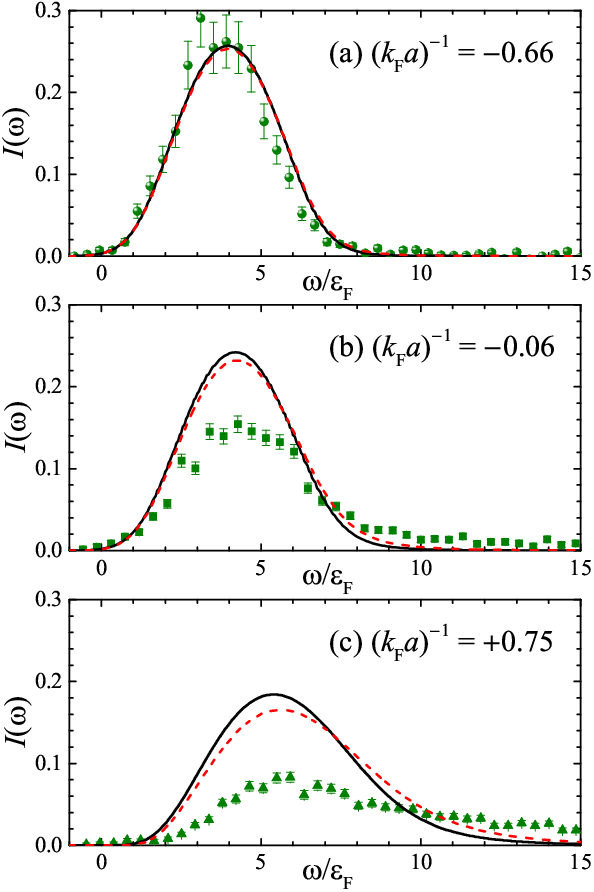}
\par\end{centering}
\caption{\label{fig:fig10_RamanExpt} The ejection Raman spectra, predicted
by the self-consistent (black solid lines) and non-self-consistent
(red dashed lines) many-body $T$-matrix theories, are compared with
the experimental data from TIIT (circles) at three different interaction
strengths \cite{Ness2020}. Here, we use the impurity density $n_{\textrm{imp}}=0.23n$
and the transferred momentum $Q=1.9k_{F}$, as in the experiment.
The temperature is set to be $T=0.2T_{F}$. In the theoretical calculations,
we do not include the inhomogeneous density profiles due to external
harmonic traps, which only lead to small, unimportant changes to the
predicted Raman spectrum \cite{Hu2022c}.}
\end{figure}

Finally, let us briefly discuss the Raman spectroscopy. In Fig. \ref{fig:fig9_RamanUnitary},
we report the Raman spectrum of a unitary Fermi polaron at two temperatures,
$T=0.2T_{F}$ and $T=0.5T_{F}$, calculated by using both $T$-matrix
theories. At both temperatures, the self-consistency treatment does
not lead to notable changes. This probably can be understood from
the fact that the Raman spectrum at a large transferred momentum $Q\sim k_{F}$
is mainly contributed by Fermi polarons at finite momentum with $k\sim k_{F}$,
for which the self-consistency in the impurity Green function becomes
less important.

In Fig. \ref{fig:fig10_RamanExpt}, we compare the predictions of
the two $T$-matrix theories with the latest measurement on Raman
spectrum from TIIT (Technion-Israel Institute of Technology) at temperature
around $0.2T_{F}$. For a weak interaction between the impurity and
the Fermi sea in Fig. \ref{fig:fig10_RamanExpt}(a), the predicted
Raman spectra from the two $T$-matrix theories are indistinguishable
at the scale of the figure. Both predictions agree well with the measured
spectrum. For large interaction strengths, near the unitary limit
(Fig. \ref{fig:fig10_RamanExpt}(b)) or on the BEC side the Feshbach
resonance (Fig. \ref{fig:fig10_RamanExpt}(c)), the two $T$-matrix
theories do predict different theoretical Raman spectrum. However,
the improvement due to the self-consistency treatment seems to be
too small, to resolve the puzzling discrepancy found earlier between
the non-self-consistent $T$-matrix theoretical results and the experimental
observations \cite{Hu2022c}.

\section{Conclusions and outlooks}

In conclusion, we have studied the spectral function of Fermi polarons
at finite temperature in three dimensions, by applying a self-consistent
many-body $T$-matrix theory. In comparison with the widely used non-self-consistent
$T$-matrix theory \cite{Combescot2007,Tajima2019,Hu2022}, we find
the introduction of the self-consistency in the impurity Green function
quantitatively changes the polaron spectral function. The changes
are mostly significant at low temperature, where the self-consistency
treatment enlarges the polaron decay rate and therefore notably broadens
the lineshape of the spectral function. We have related the enhanced
polaron decay rate to the two-particle vertex function, which describes
the in-medium molecule state created by the successive scatterings
between the impurity and the Fermi sea.

Although we believe that the molecule state is more accurately described
by the self-consistent many-body $T$-matrix theory, there is no consensus
that the self-consistency in the impurity Green function will necessarily
improve the description of Fermi polarons. Therefore, we have appealed
to the comparison of the theoretical predictions from both self-consistent
and non-self-consistent theories with the latest experimental measurements
\cite{Zan2019,Ness2020}. For the radio-frequency spectroscopy of
a unitary Fermi polarons \cite{Zan2019}, at low temperature (i.e.,
below $0.8T_{F}$) we observe that the self-consistent $T$-matrix
theory seems to explain better the experimental data. However, at
large temperature, both $T$-matrix theories fail to account for the
experimental observations. For the Raman spectroscopy of Fermi polarons
\cite{Ness2020}, the use of the self-consistent $T$-matrix theory
does not lead to too much difference. Near the Feshbach resonance,
the predictions of both $T$-matrix theories differ largely with the
experimental results at $T\sim0.2T_{F}$. The discrepancy between
$T$-matrix theories and experiments, both for radio-frequency spectroscopy
and Raman spectroscopy, suggests that we need to significantly improve
the theory of polaron spectral function, beyond the standard $T$-matrix
approximation.

Alternatively, we may consider the case of heavy Fermi polarons, where
the exact solution can be obtained by using a functional determinant
approach \cite{Schmidt2018,Wang2023AB}. The exact spectral function
of heavy polarons in a Bardeen-Cooper-Schrieffer (BCS) Fermi superfluid
was recently determined \cite{Wang2022PRL,Wang2022PRA}. It would
be interesting to calculate the spectral function of such heavy BCS
polarons with both self-consistent and non-self-consistent many-body
$T$-matrix theories and compare the approximate results with the
exact solution. Finally, it might be useful to note that, our technique
used to calculate a dressed Green function in real frequency could
be extended to investigate the spectral function of other interacting
Fermi systems, particularly a two-component spin-1/2 Fermi gas with
balanced population in each component. 
\begin{acknowledgments}
This research was supported by the Australian Research Council's (ARC)
Discovery Program, Grants Nos. DP240101590 (H.H.) and DP240100248
(X.-J.L.).
\end{acknowledgments}

\appendix
\begin{figure}
\begin{centering}
\includegraphics[width=0.45\textwidth]{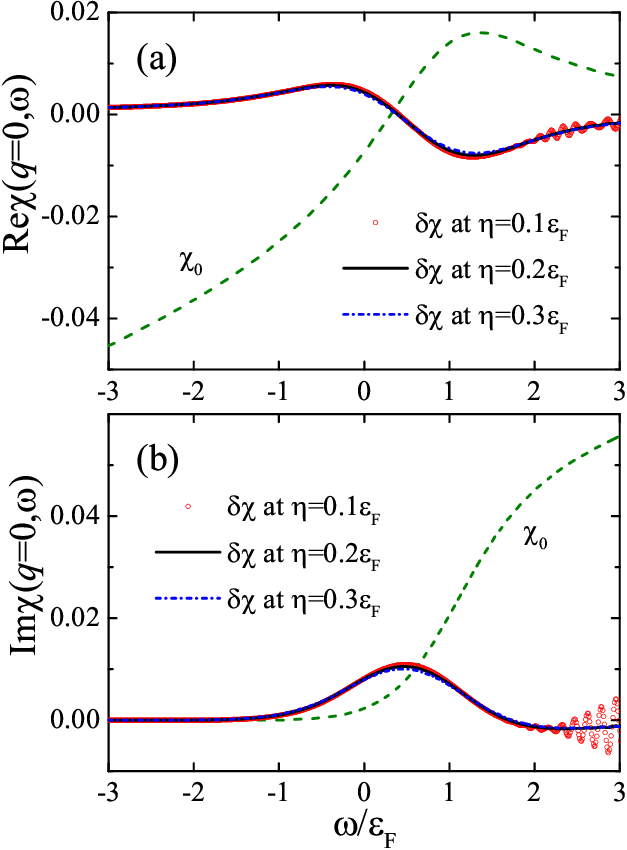}
\par\end{centering}
\caption{\label{fig:figA1_kappa} The real part and imaginary part of the pair
propagator $\chi_{0}(q,\omega)=\Gamma_{0}^{-1}(q,\omega)$ and $\delta\chi(q,\omega)$
at zero wavevector $q=0$, in arbitrary units. The red circles, black
solid lines and blue dot-dashed lines show the results with $\eta/\varepsilon_{F}=0.1$,
$0.2$ and $0.3$, respectively. We consider the unitary limit and
a temperature $T=0.2T_{F}$.}
\end{figure}

\appendix
\begin{figure}
\begin{centering}
\includegraphics[width=0.45\textwidth]{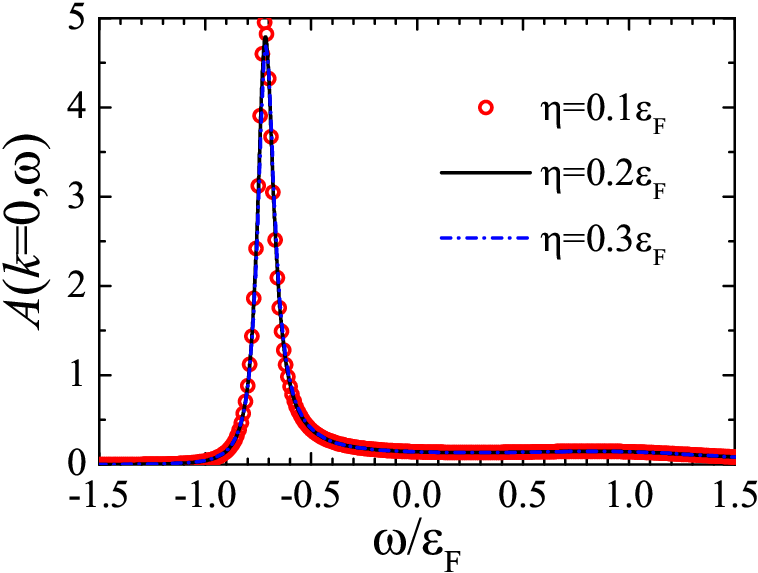}
\par\end{centering}
\caption{\label{fig:figA2_akw} The zero-momentum spectral function, in units
of $\varepsilon_{F}^{-1}$, calculated with with $\eta/\varepsilon_{F}=0.1$
(red circles), $0.2$ (black solid lines) and $0.3$ (blue dot-dashed
lines). We consider the unitary limit and a temperature $T=0.2T_{F}$.}
\end{figure}

\section{The $\eta$-dependence of the difference in vertex function}

Throughout the work, in the calculation of the difference in vertex
function $\delta\chi\equiv\Gamma^{-1}-\Gamma_{0}^{-1}$, we have added
a small imaginary part $\eta=0.2\varepsilon_{F}$ to the frequency
$\omega$, in order to remove the singularity in the integrand of
Eq. (\ref{eq:dkappa}). In Fig. \ref{fig:figA1_kappa}, we show the
dependence of $\delta\chi(q=0,\omega)$ at zero momentum on the choice
of the value $\eta$, in comparison with the dominant contribution
$\chi_{0}=\Gamma_{0}^{-1}(q=0,\omega)$. Here, as an example, we consider
the unitary limit at temperature $T=0.2T_{F}$. At the scale of $\Gamma_{0}^{-1}$,
we can barely notice the changes in $\delta\chi$ due to the use of
different values of $\eta$. We find that the use of a smaller value
of $\eta=0.1\varepsilon_{F}$ introduces an oscillation at large frequency.
This is anticipated, since the integrand needs to be more finely sampled
in our gaussian quadrature integration, which is time-consuming. Our
choice of $\eta=0.2\varepsilon_{F}$ turns to be a good balance selection,
so the numerical calculations can be carried out in an efficient and
accurate way. In Fig. \ref{fig:figA2_akw}, we also report the corresponding
spectral functions at different values of $\eta$. The three spectral
functions are indistinguishable with each others.

\begin{figure}
\begin{centering}
\includegraphics[width=0.5\textwidth]{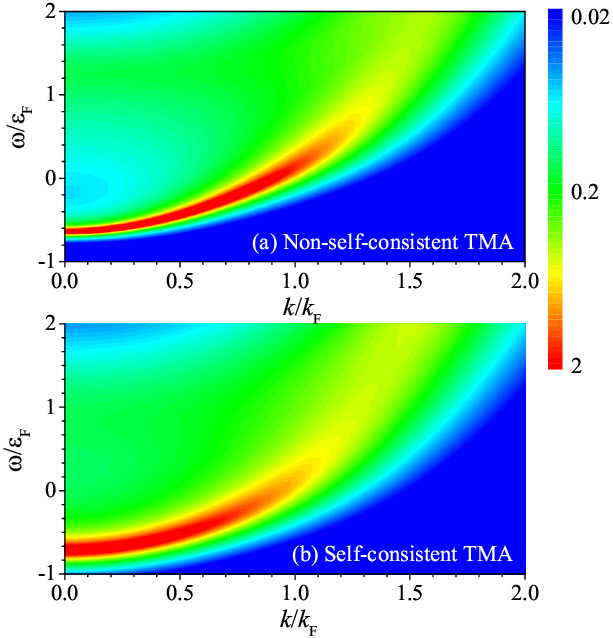}
\par\end{centering}
\caption{\label{fig:figA3_akw2dk} The contour plot of the impurity spectral
function $A(k,\omega)$ of a unitary Fermi polaron at $T=0.2T_{F}$
as functions of momentum $k$ and frequency $\omega$, predicted by
the non-self-consistent (a) and self-consistent (b) many-body $T$-matrix
theories. The two-dimensional plot is shown at a logarithmic scale
in units of $\varepsilon_{F}^{-1}$, as indicated by the color bar. }
\end{figure}

\section{Spectral function at finite momentum}

In Fig. \ref{fig:figA3_akw2dk}, we present the polaron spectral function
at finite momentum, in the form of a two-dimensional contour plot.
We consider the unitary limit with $1/(k_{F}a)=0$ and a temperature
$T=0.2T_{F}$. In comparison with the non-self-consistent $T$-matrix
results in the upper panel, we find that the self-consistency treatment
in the lower panel leads to a much broader polaron peak. As the momentum
increases, it also makes the polaron easier to dissolve.

\end{document}